\documentclass[prl,superscriptaddress,showpacs,floatfix, twocolumn]{revtex4}
\usepackage[dvips]{graphicx}
\usepackage{epsfig}
\usepackage{pst-plot}
\usepackage{bm}
\usepackage{bbm}
\usepackage{mathrsfs}
\usepackage{amsfonts}
\usepackage{amssymb}
\usepackage{lscape}

\begin{document}

\title{Complex coupled-cluster approach to an {\it ab-initio} description of 
open quantum systems}
\author{G.~Hagen}
\affiliation{Physics Division, Oak Ridge National Laboratory, P.O. Box 2008, 
Oak Ridge, Tennessee 37831, USA}  
\affiliation{Department of Physics and Astronomy, University of Tennessee, 
  Knoxville, Tennessee 37996, USA}

\author{D.J.~Dean}
\affiliation{Physics Division, Oak Ridge National Laboratory, P.O. Box 2008, 
Oak Ridge, Tennessee 37831, USA}

\author{M.~Hjorth-Jensen}
\affiliation{Department of Physics and Center of Mathematics for Applications, 
University of Oslo, N-0316 Oslo, Norway}

\author{T.~Papenbrock}
\affiliation{Department of Physics and Astronomy, University of Tennessee, 
Knoxville, Tennessee 37996, USA}
\affiliation{Physics Division, Oak Ridge National Laboratory, P.O. Box 2008, 
Oak Ridge, Tennessee 37831, USA}  

\date{\today}
\begin{abstract}
We develop {\it ab-initio} coupled-cluster theory to describe
resonant and weakly bound states along the neutron drip line. We
compute the ground states of the helium chain $^{3-10}$He within 
coupled-cluster theory in singles and doubles (CCSD) approximation. 
We employ a spherical
Gamow-Hartree-Fock basis generated from the low-momentum N$^3$LO
nucleon-nucleon interaction. This basis treats bound, resonant,
and continuum states on equal footing, and is therefore optimal for the
description of properties of drip line nuclei where continuum features
play an essential role.  Within this formalism, we present an
\emph{ab-initio} calculation of energies and decay widths 
of unstable nuclei starting from realistic interactions.

\end{abstract}
\maketitle
Exotic phenomena emerge in weakly bound and resonant many-body quantum systems. 
These phenomena include ground states that are embedded in the continuum, melting and 
reorganizing of shell structures, extreme matter clusterizations and halo densities.
These unusual features occur in many branches of
physics; as examples, we mention Fano resonances~\cite{fano} in quantum
dots~\cite{Bul01}, ultracold atom gases~\cite{Ket99}, auto-ionizing
atoms~\cite{Burg05} or molecules~\cite{Col86}, and exotic nuclei.  In
nuclear physics we find such exotic systems moving away from the valley of 
nuclear stability towards the drip lines, where
the outermost nucleons literally start to drip from the nuclei.

The theoretical description of weakly bound and unbound 
quantum many-body systems is a challenging undertaking. 
The proximity of the scattering continuum 
in these systems implies that they should be treated as 
open quantum systems where coupling with the scattering 
continuum can take place. Recent work with Gamow states employed in
Hamiltonian diagonalization  methods
\cite{witek1, roberto, michel1, hagen3, Jimmy_PRL2}
have shown that these basis states correctly depict properties associated
with open quantum systems.
This Berggren basis is composed of
bound, resonant, and (continuum) scattering  single-particle states
\cite{berggren}. 
This basis significantly
improves and facilitates the description of loosely bound systems and is 
essential in the description of unbound systems. 
However, the typically large number of discretized continuum states 
limits this approach to traditional shell-model diagonalization 
calculations where an inert core is employed. 

In this Letter, we present an {\it ab-initio} approach to open quantum
systems using a Gamow-Hartree-Fock basis derived from realistic 
interactions \cite{hagen3}. 
We employ coupled-cluster
theory~
\cite{coester, coesterkummel, kummel2, cizek, palduscizek, mihaila, cc1, cc2,
cc3} 
to solve the quantum many-body problem for the helium chain 
in this basis.  Coupled-cluster techniques computationally scale
much more gently with increasing system size, than exact
diagonalization methods, and are therefore very well suited for 
open quantum systems where the number of orbitals are typically   
orders of magnitude larger than for closed quantum systems.
Its application with Gamow basis states is
based on a non-Hermitian representation of the many-body Hamiltonian. 
This is a rather new direction
in coupled-cluster theory~\cite{Saj05}, and we report its first
successful application in nuclear theory. Other {\it ab-initio}
methods like the Green's function Monte Carlo~\cite{Pud97} or the
no-core shell model~\cite{Cau06} have previously been employed to
compute the structure of helium isotopes. 

This Letter is organized as follows. We first introduce
coupled-cluster theory, the interaction and the model space.
Second, we provide several checks to gauge the accuracy of
our approach by comparison with exact diagonalization methods.  Third,
we perform large-scale calculations of the ground states of helium
isotopes.

{\it Method and model space.} In coupled-cluster theory we make the
exponential ansatz for the exact correlated ground state,
\begin{equation}
\label{ansatz}
\vert\Psi\rangle = \exp(T) \vert \Phi_0 \rangle \ .
\end{equation}
Here $\vert \Phi_0\rangle$ is an uncorrelated reference
Slater determinant which might be either the Hartree-Fock (HF)
state or a naive filling of the oscillator single-particle basis. 
Correlations are introduced through the exponential $\exp(T) $
operating on $ |\Phi_0\rangle$. The operator $T$ is a sum of
$n$-particle--$n$-hole excitation operators $T = T_1 + T_2 + ...$ of the
form,
\begin{equation}
T_n = \sum_{a_1\ldots a_n,i_1\ldots i_n}t_{i_1\ldots i_n}^{a_1\ldots a_n}a_{a_1}^\dagger \cdots a_{a_n}^\dagger
a_{i_n}\cdots a_{i_1} \ ,
\end{equation}
where $i_1,i_2,...$ are summed over hole states and
$a_1,a_2,...$ are summed over particle states.
One obtains the algebraic equation for the excitation amplitudes
$t_{ij...}^{ab...}$ by left-projecting the similarity-transformed 
Hamiltonian with an $n$-particle--$n$-hole excited Slater
determinant giving
\begin{equation}
\label{cceq}
\langle \Phi_{ij...}^{ab...} \vert 
\left( H_N \exp(T) \right)_C \vert \Phi_0 \rangle = 0 \ ,
\end{equation}
where the Hamiltonian enters in normal-ordered form, and the subscript
$C$ indicates that only connected diagrams enter. 
We iteratively solve the non-linear set
of coupled equations~(\ref{cceq}) for the excitation amplitudes.
The solutions determine the coupled-cluster
correlation energy
\begin{equation}
\label{ccerg} 
E_{\mathrm{CC}} = \langle \Phi_0 \vert 
\left( H_N \exp(T) \right)_C \vert \Phi_0 \rangle \ .
\end{equation}
In this work, we truncate the cluster operator $T$ at the 
two-particle--two-hole level (CCSD), i.e. we approximate $T=T_1+T_2$.
We also investigate whether the 
perturbative triples correction CCSD(T) \cite{pople} improve on 
the CCSD results.

We construct our basis using the Berggren formalism
\cite{berggren} in which bound, resonant and continuum states are
treated on equal footing.  The Berggren basis is an analytic
continuation of the usual completeness relation in the complex energy
plane.  The representation of the Hamiltonian in a finite Berggren
basis is no longer Hermitian but rather complex symmetric,
and renders the coupled-cluster equations~(\ref{cceq}) 
and~(\ref{ccerg}) complex.

The nuclear Hamiltonian is given by
\begin{equation}
H = t - t_{\mathrm{CoM}} + V \ . 
\end{equation}
Here, $t$ denotes the operator of the kinetic energy, and $t_{\rm
CoM}$ is the kinetic energy of the center of mass.  The
nucleon-nucleon interaction $V$ is based on chiral effective field
theory within the N$^3$LO expansion \cite{n3lo1}. This
potential is a systematic momentum-space expansion to fourth order of 
a Lagrangian that obeys QCD symmetries. It contains 
high-momentum components and is therefore not suitable for the
limited basis sets we employ.  In order to make the calculation
feasible, we construct a low-momentum interaction $V=V_{{\rm low-}k}$
following the formalism outlined in \cite{vlowk}.
This is done by
integrating out those high-momentum modes of the chiral potential that
exceed the chosen momentum cutoff $\Lambda$.  The construction of $V_{{\rm
low-}k}$ is a renormalization group transformation and therefore
generates three-body forces and also forces of higher rank. These
forces depend on the cutoff, and only the sum of all forces is
cutoff-independent. In this work, we limit ourselves to two-body
forces. Since we are interested in helium isotopes, 
$\Lambda=1.9$fm$^{-1}$ is a {\it convenient} choice, as
the ground state expectation value of the omitted three-nucleon force
is very small for $^3$H and $^4$He \cite{nogga}. 

We build our coupled-cluster reference state from a single-particle
basis obtained through a self-consistent Gamow-HF
calculation~\cite{hagen3}.  For the helium isotopes considered in this
work, the proton separation energy is typically of the order of
$20-30$~MeV, and protons mainly occupy deeply bound $s$-orbits.
The situation is different for the neutrons where in neutron-rich 
systems the separation energy is very small.  
Furthermore, neutrons in
$p$-orbits are believed to build up the main part of surface densities.
Based on these
observations we use harmonic oscillator wave functions (with $\hbar
\omega = 20$~MeV) for the protons and for the higher partial waves
($d-g$ waves) on the neutron side.  For neutrons in $s$ and $p$
orbits, we use a complex Woods-Saxon basis where the non-resonant
continuum is defined on a triangular contour in the complex $k$-plane
(see Fig.~3 in Ref.~\cite{hagen3} for details).  Using Gauss-Legendre
quadrature, the discretization of $L^+$ has been carried out with $3$
points in the interval $(0,A)$, $4$ points in the interval $(A,B)$, and
13 points in the interval $(B,C)$. Consequently, for each of the 
$s$-$p$ partial waves on the neutron side, we have a discretized basis 
built from bound, resonant, and non-resonant continuum states.  
For all other partial waves on the
proton and neutron side, we use an oscillator basis with the energy
truncation $ N = 2n+l \leq 10 $.  This combination of complex
Woods-Saxon states for low values of angular momentum and harmonic
oscillator states for higher values of angular momentum captures the
relevant physics and keeps the total size of the single-particle basis
manageable.  We find good convergence of the HF energy with
respect to the number of integration points and size of our single-particle 
model space. 

{\it Accuracy of the Coupled-Cluster method.} Weakly bound and resonant nuclei present a
double challenge to the coupled-cluster method. First, some of 
the considered helium isotopes have open-shell character.  
Such systems are more difficult to
describe within single-reference coupled-cluster methods.  Second,
particle-unstable nuclei like $^{5,7}$He have resonant ground
states. Here, the physical ground state is not the ground state of the
model space we employ since scattering states might have lower
energies. We develop a procedure which allows one to identify the  
physical state on the many-particle energy surface. 
Both problems are addressed in what follows.

To study the accuracy for open-shell nuclei, we compare the CCSD
energies of $^{3-6}$He with exact results obtained through
diagonalization. We restrict ourselves to Hamiltonians
represented in a finite oscillator space,
and thereby separate open-shell aspects from properties related to
open systems. The exact diagonalization is only possible in a
relatively small model space consisting of $s$, $p$, and $d$ states up
to the $4s3p1d$ oscillator states.  The results are presented in
Table~\ref{tab:exact_ccsd}. The CCSD calculations use a reference
Slater determinant built from a spherical oscillator (OSC) basis, from a
spherical spin-restricted HF basis (RHF), and from a semi-canonical HF 
basis in which 
the Fock-matrix is diagonal in the hole/hole and particle/particle subspaces 
(SC-RHF).
The basis sets are spherically symmetric, 
and there is a freedom in defining a reference Slater
determinant for open-shell nuclei. For a nucleus with known spin $J$,
we define our reference state such that its total spin projection is
maximal. Furthermore, the orbits with largest absolute value of the
spin projection $m_j$ are filled first. For example, for $^6$He we
place the two outermost neutrons in the $m_j = 3/2, -3/2$ orbitals for
the ground state calculation. 
In Table~\ref{tab:exact_ccsd} we compare the
results from diagonalization with the CCSD results and with
triples-corrected results (CCSD(T)). The perturbative triples corrections 
are calculated using converged $T_1$ and $T_2$ amplitudes.
For $^{3-5}$He the
CCSD results differ by not more than 500 keV from the exact
results. Triples corrections improve this deviation to 200 keV (or less). 
For the open-shell nucleus $^6$He, the CCSD results differ by 1.7
MeV
from the exact result, including triples correction 
the error decreases to 200 keV.

\begin{table}[hbtp]
    \begin{tabular}{lllll}
      \hline
      \multicolumn{1}{c}{Method} & \multicolumn{1}{c}{ $^3$He}
      & \multicolumn{1}{c}{ $^4$He}
      & \multicolumn{1}{c}{ $^5$He} 
      & \multicolumn{1}{c}{ $^6$He} \\
      \hline
      CCSD (OSC)      & -6.21  &  -26.19 &   -21.53 & -20.96  \\
      CCSD (RHF)      & -6.10  &  -26.06 &   -21.55 & -20.99  \\
      CCSD (SC-RHF)   & -6.11  &  -26.06 &   -21.55 & -21.04  \\ 
      CCSD(T) (OSC)   & -6.40  &  -26.30 &   -21.91 & -22.83   \\
      CCSD(T) (RHF)   & -6.35  &  -26.24 &   -21.90 & -22.56   \\
      CCSD(T) (SC-RHF) & -6.34  &  -26.24   &  -21.91   & -22.62    \\
      Exact            & -6.45  &  -26.3  &   -22.1  & -22.7   \\
      \hline
    \end{tabular}
  \caption{Comparison of CCSD results and triples-corrected CCSD(T)
results with exact calculations for the ground states of helium
isotopes. The energies $E$ are given in MeV, and the results are displayed
for different basis sets as described in the text.}
    \label{tab:exact_ccsd} 
\end{table}
Using different basis sets, the CCSD(T) results for 
$^{3-5}$He do not vary by more than $\sim 60$keV, indicating improved convergence with CCSD(T). 
However, for $^6$He the CCSD(T) results vary by $\sim 300$keV for the different basis sets used.  
This indicates that
the perturbative triples correction CCSD(T) is not tenuous for the 
nucleus $^6$He, and that the triples clusters have to be treated more accurately
for truly open-shell nuclei \cite{piotr2,noga2}. 

To study the accuracy of CCSD for particle-unstable nuclei, we
consider the problem of $^7$He (using a $^4$He core) and compare 
with exact diagonalizations. Recall that the resonant 
state is embedded in a (quasi) continuum of scattering states. 
Thus, one must construct a procedure to identify it. Within CCSD, 
we use a reference state built from bound and resonant single-particle
orbitals. Therefore, the reference state is a localized 
state in the Gamow-HF basis and the 
CCSD correlations are built upon it.  Our model space 
for $^7$He consists of nine
$p_{3/2}$ orbitals above the $^4$He core. The exact diagonalization
yields a resonant state at energy $E=2.37$MeV and width
$\Gamma=0.23$MeV, our CCSD result deviates from this result by
less than 10 keV. 
We also checked that 
the results reported in this Letter show good convergence with respect to 
the number of discretization points of the contour $L^+$, 
and with respect to changes of the oscillator frequency 
of the basis states we employ. 
We estimate the error due to the limited discretization,
to be within $100$ keV for the real part and $20$keV for the imaginary part of 
the energy.

{\it Results.} We now turn to large-scale CCSD calculations for
$^{3-10}$He isotopes.  Table~\ref{tab:he3_10_conv} presents the
converged CCSD ground state energies for the $^{3-10}$He isotopes for
increasing number of partial waves in our single-particle basis. Here,
$s$-$p$ refers to a $5s5p$ proton and $20s20p$ neutron space; $s$-$d$
refers to a $5s5p5d$ proton and $20s20p5d$ neutron space; $s$-$f$ refers
to a $5s5p5d4f$ proton and $20s20p5d4f$ neutron space; finally
$s$-$g$ refers to a $5s5p5d4f4g$ proton space and a $20s20p5d4f4g$
neutron space, respectively.  For our $s$-$g$ calculation we have a
total of 556 single-particle states. The computed widths and lifetimes of the helium isotopes 
are in semi-quantitative agreement with experiment. Our CCSD calculations
correctly depict that  $^5$He and $^7$He are
unbound while $^8$He is bound  in their ground states. 
At the CCSD level, $^6$He is nearly bound in its ground state.
We found that the perturbative triples correction (T) to the ground state 
of $^6$He does not improve on the CCSD results. 
We also found that the triples correction differs considerably using
approximate or fully converged $T_1$ and $T_2$ amplitudes. 
This is contrary to what is typically found in quantum chemistry, and suggests
that a perturbative treatment becomes invalid.
It   might be
that the HF state is not a good starting point for a perturbative expansion and/or 
that the high density of continuum states makes perturbation theory break down.

Our CCSD calculations show convergence with respect to the
single-particle basis size. For example, $^5$He changes by only 300~keV (to
-24.87~MeV) when we add $g$-orbitals.  An extrapolation of the $^5$He
result to an infinite space using $E_{N} =
\alpha\exp(-N/Nt)+E_\infty$, where $N$ represents the space size, and
$\alpha$ and $Nt$ are fit parameters, yields
$E_\infty=-24.89\pm0.01$~MeV. A similar fit for the $^8$He data yields
$E_\infty=-26.90\pm0.03$~MeV which is about $700$~keV below the
$s$--$g$ calculations. Thus, the largest calculations we are
performing appear to have good convergence with the number of basis
states,
and one is able to perform a simple exponential fit to obtain full space results
with estimates on the extrapolation error.  We note that the actual
masses have a familiar pattern (based on $AV18$ results \cite{av18})
of underbinding as one increases the neutron number. As was found in
GFMC calculations, this underbinding should  mainly be overcome by the
inclusion of three-body forces.
\begin{table*}[htbp]
\begin{ruledtabular}
  \begin{tabular}{llllllllrllllllll}
      \multicolumn{1}{c}{} & \multicolumn{2}{c}{ $^3$He}
      & \multicolumn{2}{c}{ $^4$He}
      & \multicolumn{2}{c}{ $^5$He}
      & \multicolumn{2}{c}{ $^6$He} 
      & \multicolumn{2}{c}{ $^7$He}
      & \multicolumn{2}{c}{ $^8$He} 
      & \multicolumn{2}{c}{ $^9$He} 
      & \multicolumn{2}{c}{ $^{10}$He} \\
      \hline
      \multicolumn{1}{c}{ $lj$}&
      \multicolumn{1}{c}{Re[E]}&\multicolumn{1}{c}{Im[E]} &
      \multicolumn{1}{c}{Re[E]}&\multicolumn{1}{c}{Im[E]} &
      \multicolumn{1}{c}{Re[E]}&\multicolumn{1}{c}{Im[E]} &
      \multicolumn{1}{c}{Re[E]}&\multicolumn{1}{c}{Im[E]} &
      \multicolumn{1}{c}{Re[E]}&\multicolumn{1}{c}{Im[E]} &
      \multicolumn{1}{c}{Re[E]}&\multicolumn{1}{c}{Im[E]} &
      \multicolumn{1}{c}{Re[E]}&\multicolumn{1}{c}{Im[E]} & 
      \multicolumn{1}{c}{Re[E]}&\multicolumn{1}{c}{Im[E]} \\
      \hline
      $s-p$ & -4.94 & 0.00 & -24.97 & 0.00 & -20.08 & -0.54 & -19.03 & -0.18  
      &  -17.02 & -0.24 &  -16.97  & -0.00 & -15.28 & -0.40 & -13.82 & -0.12  \\
      $s-d $& -6.42 & 0.00 & -26.58 & 0.00 & -23.56 & -0.22 & -23.26 & -0.09 
      &  -22.19 & -0.12 &  -22.91  & -0.00 & -21.34 & -0.15 & -20.60 & -0.02  \\
      $s-f$ & -6.81 & 0.00 & -27.27  & 0.00 & -24.56 & -0.17 & -24.69 & -0.07      
      & -24.13 & -0.11   &  -25.28  &-0.00 & -23.96 & -0.06 & -23.72 & -0.00   \\
      $s-g$ & -6.91 & 0.00 & -27.35 & 0.00 & -24.87 & -0.16 & -25.16 & -0.06    
      & -24.83  & -0.09  &  -26.26  & -0.00 & -25.09 &  -0.03  & -24.77  & -0.00  \\
      \hline 
      Expt. & -7.72 & 0.00 & -28.30 & 0.00 & -27.41 & -0.33(2) & -29.27 & 0.00 
      & -28.83  & -0.08(2) &  -31.41  &  0.00 & -30.14 &  -0.05(3) &  -30.34 & -0.09(6) \\
    \end{tabular}
    \end{ruledtabular}
  \caption{    \label{tab:he3_10_conv} 
    CCSD calculation of the $^{3-10}$He ground states with the
    low-momentum N$^3$LO nucleon-nucleon interaction for increasing
    number partial waves.  The energies $E$ are given in MeV for both
    real and imaginary parts.  Experimental data are from
    Ref.~\cite{audi}. }
\end{table*}
We also compute the ground state expectation value of $J^2$.  In the
case of exact or variationally determined wave functions, the
expectation value of an operator $O$ can be evaluated via the
Hellmann-Feynman theorem, $\left. {dE \over d\lambda }
\right|_{\lambda = 0} = \langle \psi(0) \vert O \vert \psi(0) \rangle$
by adding the small perturbation $\lambda O$ to the Hamiltonian.
Though coupled-cluster theory is not a variational theory, the
Hellmann-Feynman theorem is {\it effectively} fulfilled provided the
ground state is determined with sufficient accuracy \cite{noga}.  We
find that the spins of all nuclei are well reproduced (to about one
part in 1000) compared with experimental values, except for $^6$He
where the CCSD result is $J=0.6$. 
It seems that a full CCSDT
calculation would be needed to improve this expectation value. 

In summary, we applied coupled-cluster theory for the {\it ab-initio}
description of loosely bound and unbound nuclei. This is the first microscopic
calculation that computes lifetimes of unstable nuclei 
from realistic nucleon-nucleon interactions.
Using a renormalized interaction of the low-momentum type, basic properties of the helium
chain are reproduced, i.e.  $^{5,7}$He unbound, $^{8}$He
bound and $^6$He nearly bound at the CCSD level.  
The decay widths of unbound nuclei are in semiquantitative
agreement with experimental data. For small model spaces, we could
verify that the employed CCSD approximation agrees well with results
from exact diagonalizations, and that CCSD(T) corrections improve
our open-shell results. However, our CCSD(T) results for $^6$He 
indicates that triples corrections cannot be treated perturbatively
in the case of systems with a truly open-shell character. Different schemes for
including triples corrections in these systems will be 
investigated in the future. 

We acknowledge discussions with Nicolas
Michel and Jimmy Rotureau.
This work was supported in part by the U.S.~Department of
Energy under Contract Nos.~DE-AC05-00OR22725 (Oak Ridge National
Laboratory), DE-FG02-96ER40963 (University of Tennessee),
DE-FG05-87ER40361 (Joint Institute for Heavy Ion Research), and by the
Research Council of Norway (Supercomputing grant NN2977K). Computational
resources were provided by the Oak Ridge Leadership Class Computing
Facility and the National Energy Research Scientific Computing
Facility.


\end{document}